\documentclass[aps,prl,twocolumn,preprintnumbers,superscriptaddress,
              showpacs,nofootinbib]{revtex4}
\newcommand{\PRE}[1]{}       

\usepackage{bm} \usepackage{epsfig}

\newcommand{\OmegaDM}{\Omega_{\text{DM}}}

\newcommand{\kev}{\text{keV}} 
\newcommand{\mev}{\text{MeV}}
\newcommand{\gev}{\text{GeV}} 
\newcommand{\tev}{\text{TeV}}

\newcommand{\s}{\text{s}}
 
\newcommand{\Mpc}{\text{Mpc}} 
\newcommand{\etal}{{\em et al.}}

\newcommand{\eqref}[1]{Eq.~(\ref{#1})}

\newcommand{\figref}[1]{Fig.~\ref{fig:#1}}

\newcommand{\mchi}{m_{\chi}}
\newcommand{\mgravitino}{m_{\gravitino}}

\newcommand{\gravitino}{\tilde{G}}

\newcommand{\mmess}{M_{\text{m}}}
\newcommand{\lmess}{\Lambda}
\newcommand{\nmess}{N_{\text{m}}}
\newcommand{\signmu}{\text{sign}(\mu)}
\newcommand{\Omegachi}{\Omega_{\chi}}
\newcommand{\lambdafs}{\lambda_{\text{FS}}}

\begin{document}

\preprint{UCI-TR-2007-37}
\preprint{CLNS 07/2006}

\title{
\PRE{\vspace*{1.5in}}
Goldilocks Supersymmetry:
Simultaneous Solution to the Dark Matter and Flavor Problems of 
Supersymmetry
\PRE{\vspace*{0.3in}}
}

\author{Jonathan L.~Feng}
\affiliation{Department of Physics and Astronomy, University of
California, Irvine, CA 92697, USA
\PRE{\vspace*{.5in}}
}

\author{Bryan T.~Smith}
\affiliation{Department of Physics and Astronomy, University of
California, Irvine, CA 92697, USA
\PRE{\vspace*{.5in}}
}

\author{Fumihiro Takayama%
\PRE{\vspace*{.2in}}
} 
\affiliation{Institute for High Energy Phenomenology, Cornell
University, Ithaca, NY 14853, USA
\PRE{\vspace*{.5in}}
}


\begin{abstract}
\PRE{\vspace*{.3in}} Neutralino dark matter is well motivated, but
also suffers from two shortcomings: it requires gravity-mediated
supersymmetry breaking, which generically violates flavor constraints,
and its thermal relic density $\Omega$ is typically too large.  We
propose a simple solution to both problems: neutralinos freezeout with
$\Omega \sim 10-100$, but then decay to $\sim 1~\gev$ gravitinos,
which are simultaneously light enough to satisfy flavor constraints
and heavy enough to be all of dark matter.  This scenario is naturally
realized in high-scale gauge-mediation models, ameliorates small scale
structure problems, and implies that ``cosmologically excluded''
models may, in fact, be cosmologically preferred.
\end{abstract}


\maketitle

Supersymmetric extensions of the standard model of particle physics
are among the prime candidates for new microphysics.  Among their many
virtues, supersymmetric models naturally predict new particles that
are candidates for dark matter.  The most well studied of these are
thermal relic neutralinos~\cite{Goldberg:1983nd}, superpartners of the
Higgs and electroweak gauge bosons.  The thermal relic density of
neutralinos is dependent on unknown supersymmetry parameters.
However, order-of-magnitude estimates yield relic densities that are
consistent with~\cite{Spergel:2006hy}
\begin{equation}
\OmegaDM h^2 = 0.1050^{+0.0041}_{-0.0040} \ \ (1\sigma) \ ,
\label{omega}
\end{equation}
where $\OmegaDM$ is the observed energy density of non-baryonic dark
matter in units of the critical density, and $h \simeq 0.73$ is the
normalized Hubble parameter.  This remarkable fact has not only
motivated supersymmetry, but has also focused attention on
``cosmologically preferred'' models, in which the neutralino thermal
relic density is exactly that required for dark matter.  Such studies
have implications for a large range of experiments, from direct and
indirect dark matter searches to those at the Large Hadron Collider
(LHC) at CERN.

The neutralino dark matter scenario is not without its blemishes,
however.  First, for the neutralino to be stable, it must be the
lightest supersymmetric particle (LSP). In particular, it must be
lighter than the gravitino.  This requires gravity-mediated
supersymmetry breaking models, in which low energy bounds on flavor
and CP violation are generically violated by several orders of
magnitude.  Gauge-mediated supersymmetry breaking (GMSB)
models~\cite{Dine:1981za} elegantly avoid these constraints, but such
models have gravitino LSPs and so are incompatible with neutralino
dark matter.

Second, although general arguments imply that the neutralino thermal
relic density is of the right order of magnitude, in concrete models,
it is often too large: Neutralinos are Majorana fermions, and so
annihilation to quarks and leptons is $P$-wave suppressed.  In
addition, gauge coupling unification and radiative electroweak
symmetry breaking typically imply Bino-like neutralinos, which
suppresses annihilation to gauge and Higgs bosons.  These effects
together enhance relic densities to values that may far exceed those
given in \eqref{omega}.

These two shortcomings of neutralino dark matter are usually
considered unrelated and addressed separately.  One may, for example,
consider gravity-mediated scenarios, such as minimal supergravity,
where low energy constraints are satisfied by unification assumptions.
One then further focuses on special regions of parameter space in
which the neutralino relic density is reduced to acceptable levels
through, for example, resonant annihilation~\cite{Griest:1990kh}, stau
co-annihilation~\cite{Ellis:1998kh}, or significant Bino-Higgsino
mixing~\cite{Feng:2000gh}.  Alternatively, one may simply abandon the
hope that the order-of-magnitude correctness of the neutralino thermal
relic density is a significant lead in the hunt for dark matter and
explore other mechanisms for dark matter production.  For example, one
may consider GMSB models with thermally produced
gravitinos~\cite{Pagels:1981ke}. (Note, however, that recent
Lyman-$\alpha$ constraints requiring $m_{\gravitino} \ge
2~\kev$~\cite{Viel:2006kd} imply that the gravitino thermal relic
density $\Omega_{\gravitino}^{\text{th}} h^2 \approx 1.2 \,
(m_{\gravitino} / \kev)$ must be significantly diluted through late
entropy production~\cite{Baltz:2001rq} for this possibility to be
viable.)  More recently, GMSB-like models with gravitino dark matter
produced by late decaying gauge singlets have also been
proposed~\cite{Ibe:2006rc}.

In this work, we consider the possibility that the two shortcomings
described above are not separate issues, but are in fact pointing to a
single resolution.  We propose that neutralinos do, in fact, freezeout
with very large densities.  However, they then decay to gravitinos,
which are light enough to accommodate the GMSB solution to the flavor
and CP problems, but heavy enough to be all of dark matter.  In
analogy to Goldilocks planets, which have temperatures that lie within
the narrow window required to support life, these supersymmetric
models have gravitino masses in the narrow window required to satisfy
both particle physics and cosmological constrants, and so we call this
``Goldilocks Supersymmetry.''

The essential features of this scenario may be illustrated by simple
scaling arguments.  Consider models in which there are two mass
scales: the scale of the standard model superpartner masses
$\tilde{m}$, and the gravitino mass $\mgravitino$.  The freezeout
density of neutralinos is inversely proportional to the neutralino
annihilation cross section, and so by dimensional analysis,
$\Omegachi h^2 \sim \langle \sigma v \rangle ^{-1} \sim
\tilde{m}^2$.  The gravitino relic density is therefore
$\Omega_{\gravitino}h^2 = (\mgravitino/\tilde{m} ) \Omegachi h^2
\sim \mgravitino \tilde{m}$.  At the same time, a natural solution to
the supersymmetric flavor and CP problems requires $\mgravitino \ll
\tilde{m}$.  We find, then, that we can always make
$\Omega_{\gravitino}$ large enough to explain dark matter by raising
$\mgravitino$ and $\tilde{m}$ together with their ratio fixed.  The
essential question, then, is whether the scenario may be realized with
$\tilde{m} \alt \tev$, as required for a natural solution to the gauge
hierarchy problem, and whether it passes all other particle physics
and astrophysical constraints.

To analyze this question concretely, we consider the example of
minimal GMSB models~\cite{Dimopoulos:1996yq}.  Such models are
specified by the 4+1 parameters $\mmess$, $\lmess$, $\nmess$,
$\tan\beta$, and $\signmu$, where $\mmess$ is the messenger mass,
$\lmess = F / \mmess$, where $F$ is the supersymmetry breaking scale
in the messenger sector, $\nmess$ is the number of ${\bf 5}+{\bf
\overline{5}}$ messenger pairs, $\tan\beta = \langle H_u^0 \rangle /
\langle H_d^0 \rangle$, and $\mu$ is the supersymmetric Higgsino mass.
In terms of these parameters, the gauge-mediated contributions to
squark and slepton masses are
\begin{eqnarray}
  m_{\tilde{f}}^2(\mmess) = 2\nmess \, \lmess^2 \sum_{i=1}^{3}
  C^f_i \left[\frac{g_i^2(\mmess)}{16\pi^2}\right]^{2} \ ,
\label{m_sf}
\end{eqnarray}
where $C^f_1=\frac{5}{3}Y^2$, with hypercharge $Y=Q-T_3$, and
$C^f_i=0$ for gauge singlets, $\frac{3}{4}$ for SU(2)$_L$ doublets,
and $\frac{4}{3}$ for SU(3)$_C$ triplets.  The gaugino masses are
\begin{eqnarray}
M_i(\mmess) = \nmess \, \lmess\, c_i \frac{g_i^2(\mmess)}{16\pi^2} \, ,
\label{m_gaugino}
\end{eqnarray} 
where $i=1,2,3$ for the U(1)$_Y$, SU(2)$_L$, and SU(3)$_C$ groups,
$c_1=\frac{5}{3}$, and $c_2=c_3=1$.  As indicated, these masses are
generated at the energy scale $\mmess$.  We determine physical masses
through renormalization group evolution to the weak scale and
radiative electroweak symmetry breaking with SoftSUSY
2.0~\cite{Allanach:2001kg}.

In addition to the gauge-mediated masses, there are gravity-mediated
contributions.  These generate the gravitino mass $\mgravitino =
\frac{F_0}{\sqrt{3} M_*}$, where $F_0$ is the total supersymmetry
breaking scale and $M_* \simeq 2.4 \times 10^{18}~\gev$ is the reduced
Planck mass.  Because $F_0$ receives contributions from all
supersymmetry breaking $F$-terms, $F_0 \ge F$. For direct gauge
mediation, $F_0 \sim F$, but this is model-dependent.  Here, we assume
$F_0 = F$, and so
\begin{equation}
m_{\gravitino} = \frac{F}{\sqrt{3} M_*} 
= \frac{\mmess \lmess}{\sqrt{3} M_*} \ .
\label{mgravitino}
\end{equation}
Our results are not changed significantly for $F_0 > F$.

Gravity-mediation also generates flavor- and CP-violating squark and
slepton mass parameters $(m_{ij}^f)_{AB}$, where $i, j = 1, 2, 3$
label generation, $A, B = L, R$ label chirality, and $f = l, u, d$.
The chirality-preserving parameters are naturally $\sim
m_{\gravitino}$; for concreteness, we assume $|(m_{ij}^f)_{LL}| =
|(m_{ij}^f)_{RR}| = m_{\gravitino}$.  The chirality-violating masses
require the breaking of electroweak gauge symmetry (and possibly
horizontal symmetries); we assume $|(m_{ij}^f)_{LR}| \alt
\lambda_{ij}^f m_{\gravitino}$, where the $\lambda_{ij}^f$ are Yukawa
couplings.  Finally, we assume ${\cal O}(1)$ CP-violating phases for
both the gravity- and gauge-mediated masses, as detailed below.

Given these assumptions, the most stringent constraints are the
flavor-changing observables $\Delta m_K$ and $\epsilon_K$, and the
CP-violating, but flavor-preserving, electron and neutron electric
dipole moments (EDMs)~\cite{Yao:2006px,Regan:2002ta,Baker:2006ts}:
\begin{eqnarray}
\Delta m_K^{\text{SUSY}} &<& 3.5~\times 10^{-12}~\mev \\
\epsilon_K^{\text{SUSY}} &<& 2.3 \times 10^{-3} \\
d_e &<& 1.6 \times 10^{-27} \ e~\text{cm} \\
d_n &<& 2.9 \times 10^{-26} \ e~\text{cm} \ .
\end{eqnarray}
In the mass insertion approximation, these constrain
$(\delta_{ij}^f)_{AB} \equiv (m_{ij}^f)_{AB} / \bar{m}_{\tilde{f}}$,
where $\bar{m}_{\tilde{f}}$ is an average $\tilde{f}$ mass.  The
leading constraints are from $\Delta m_K$ on $\text{Re} \left[
(\delta_{12}^d)_{LL} (\delta_{12}^d)_{RR} \right]$, from $\epsilon_K$
on $\text{Im} \left[ (\delta_{12}^d)_{LL} (\delta_{12}^d)_{RR}
\right]$, and from the EDMs on the gauge-mediated masses.  

The supersymmetric contributions to the kaon observables are $\Delta
m_K^{\text{SUSY}} = \text{Re} (M)$ and $\epsilon_K^{\text{SUSY}} =
\text{Im} (M) / (\sqrt{8} \, \Delta m_K^{\text{exp}})$, with $M$ as
given in Ref.~\cite{Gabbiani:1996hi}.  For concreteness, we choose the
$\delta$ phases to maximize the supersymmetric contribution for each
kaon observable.  The constraints from $\Delta m_K$ and $\epsilon_K$
are therefore not simultaneously applicable, but the most stringent
constraint smoothly interpolates between these as the phase varies.
For the EDMs, we first use micrOMEGAs 1.3.7~\cite{Belanger:2004yn} to
determine the supersymmetric contribution to $a_{\mu}$, the anomalous
magnetic moment of the muon.  The EDMs are, then, $d_e = \frac{m_e}{2
m_{\mu}^2} a_{\mu} \tan \theta_{\text{CP}}$ and $d_n = \frac{1}{3} (4
d_d + d_u)$, where $d_d$ and $d_u$ are determined from $d_e$ with
$\alpha \to \alpha_s$, $M_1 \to M_3$, $m_{\tilde{l}} \to m_{\tilde{d},
\tilde{u}}$, and the introduction of appropriate color
factors~\cite{Gabbiani:1996hi}.  We set $\tan \theta_{\text{CP}} = 1$
in the EDMs. Note that the EDMs may be suppressed, depending, for
example, on the origin of the $\mu$ and $B$ parameters.

\begin{figure}
\resizebox{3.45 in}{!}{
\includegraphics{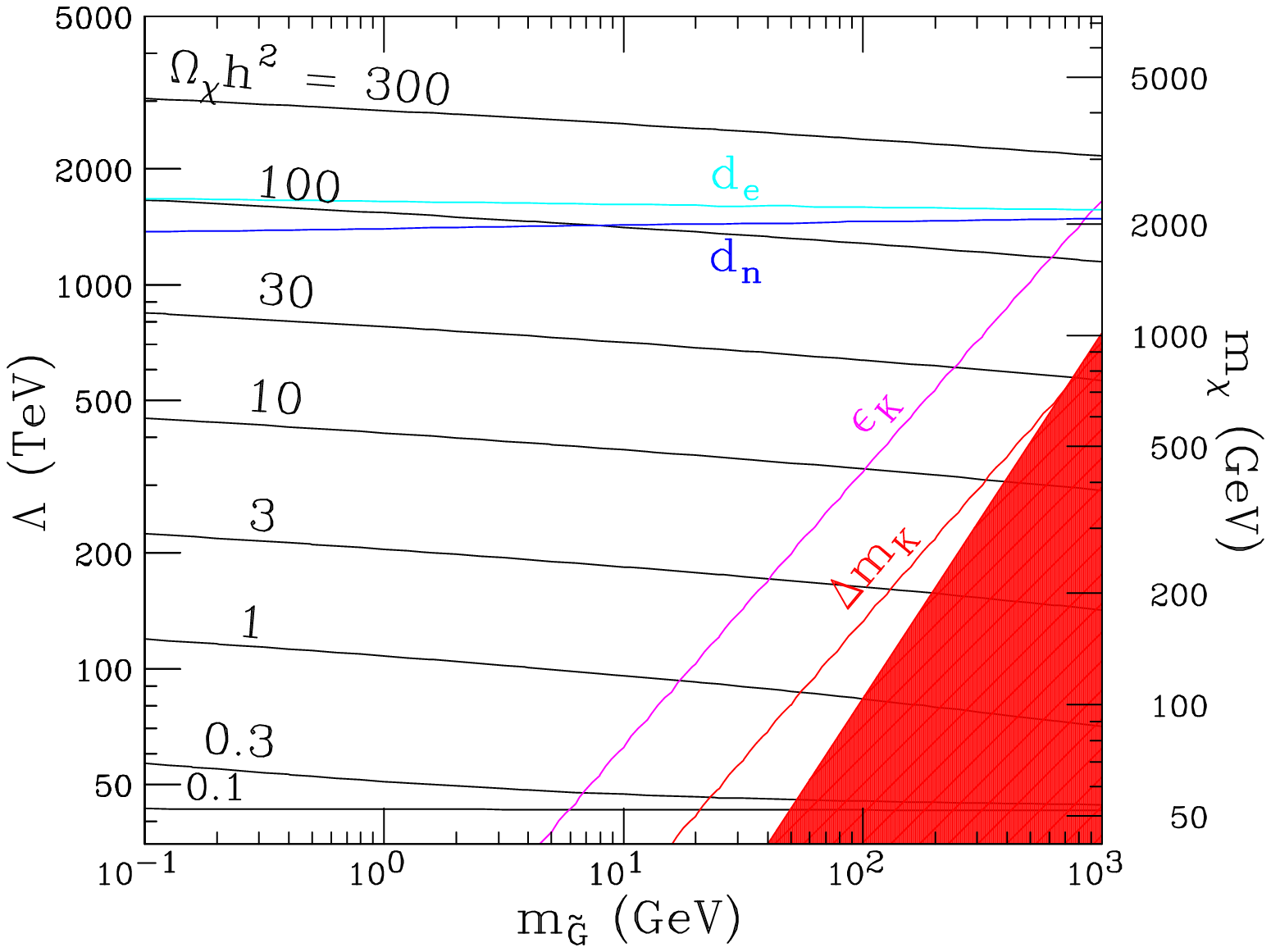}
}
\caption{Neutralino thermal relic density $\Omega_\chi h^2$ in the
$(\mgravitino, \lmess)$ plane, for $\nmess = 1$, $\tan\beta = 10$,
$\mu > 0$ and $m_t = 175~\gev$.  The right-hand axis gives the
neutralino mass $m_{\chi} \approx 1.3 \times 10^{-3} \Lambda$.
Regions to the right of the $\epsilon_K$ and $\Delta m_K$ contours and
below the $d_e$ and $d_n$ contours are disfavored.  The neutralino is
the LSP in the shaded region.
\label{fig:omegachi} 
}
\end{figure}

The resulting constraints are given in \figref{omegachi}.  The
observables $\Delta m_K$ and $\epsilon_K$ require $\mgravitino \alt
30~\gev \ (500~\gev)$ for neutralino mass $\mchi \sim 100~\gev \
(1~\tev)$.  In contrast, the EDMs are insensitive to $m_{\gravitino}$,
since they do not rely on gravity-mediated contributions.  They are
found to require $\mchi \agt 1~\tev$, in agreement with earlier
work~\cite{Moroi:1998km}.  These results are, of course, subject to
the assumptions we have made.  However, they imply that in any model
in which gravity-mediated contributions are at their natural scale and
all mass parameters have ${\cal O}(1)$ phases, the standard model
superpartners must be heavy, and the LSP is the gravitino, not the
neutralino.

For $\nmess = 1$, the lightest standard model superpartner is the
lightest neutralino $\chi$ throughout parameter space.  In
\figref{omegachi} we also show the freezeout density $\Omegachi h^2$,
that is, the relic density if neutralinos were stable, determined
using micrOMEGAs~\cite{Belanger:2004yn}.  These results illustrate the
difficulties for neutralino dark matter.  At the weak scale, typically
$\mu, M_2 > M_1$, and $\chi$ is Bino-like. Its annihilation is
therefore suppressed for the reasons noted above.  For $m_\chi =
100~\gev$, $\Omegachi h^2 \sim 1$ is already far too large, and for
the heavier superpartner masses favored by the EDM constraints, it
grows to values of $\sim 10-100$.

\begin{figure}
\resizebox{3.45 in}{!}{
\includegraphics{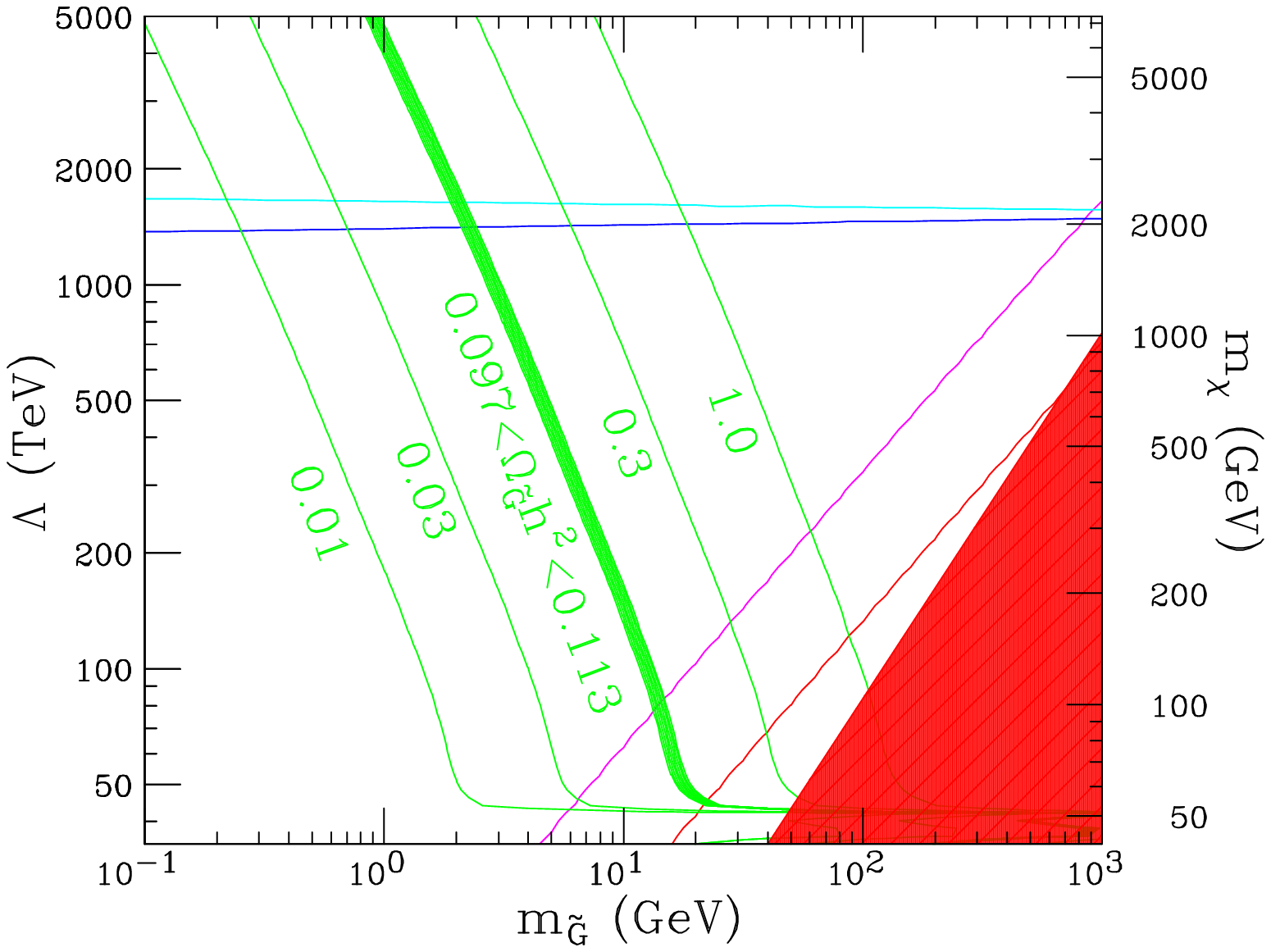}
}
\caption{Contours of $\Omega_{\gravitino} h^2$ in the $(\mgravitino,
\lmess)$ plane,.  The thick contour is the 2$\sigma$ allowed region.
Low-energy constraints and fixed GMSB parameters are as in
\figref{omegachi}.
\label{fig:omegagrav} 
}
\end{figure}

In the scenario proposed here, however, neutralinos are not stable,
but decay to gravitinos.  The resulting gravitino relic density is
given in \figref{omegagrav}.  In the dark green shaded region,
$\Omega_{\gravitino} h^2$ is in the range required to account for all
of non-baryonic dark matter.  We see that parts of this shaded region
are consistent with low energy flavor and CP constraints.  In this
scenario, very large neutralino freezeout densities are a virtue, not
a problem, as they allow light gravitinos to have the required relic
density, despite the significant dilution factor $m_{\gravitino} /
m_{\chi}$.  In this simple example of minimal GMSB, the Goldilocks
window, in which both relic density and low energy constraints are
satisfied, has $m_{\gravitino} \sim 1 - 10~\gev$.

\begin{figure}
\resizebox{3.45 in}{!}{
\includegraphics{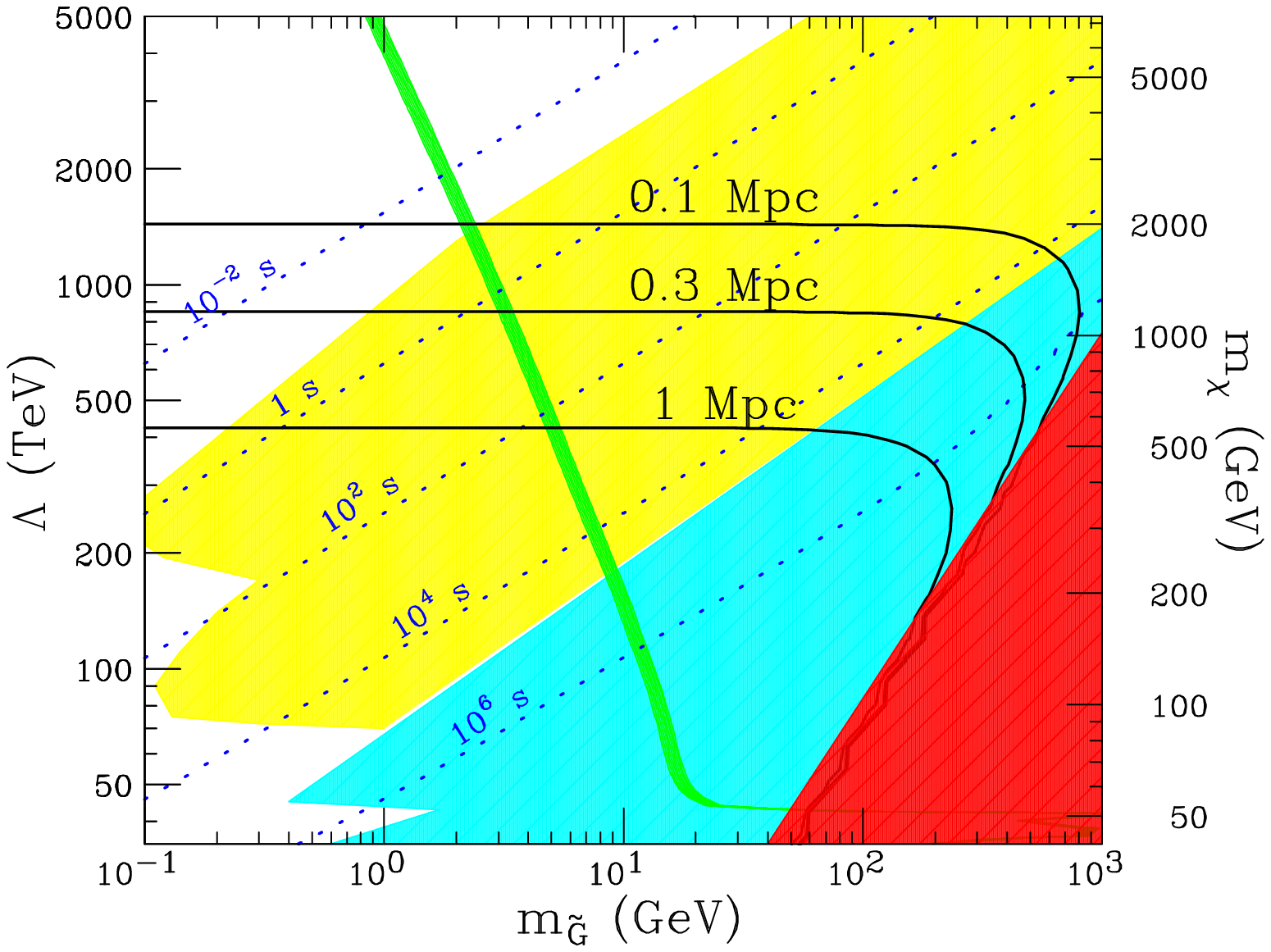}
}
\caption{Contours of $\lambdafs$ (solid) and lifetime $\tau(\chi \to
\gravitino)$ (dotted) in the $(m_{\gravitino}, \lmess)$ plane, for
$\nmess = 1$, $\tan\beta = 10$, $\mu > 0$, and top quark mass $m_t =
175~\gev$. In the light yellow (medium blue) shaded region, hadronic
(electromagnetic) showers from $\chi$ decays produce discrepancies
with BBN observations.  The band with the correct
$\Omega_{\gravitino}h^2$ is as in \figref{omegagrav}, and the
neutralino LSP region and fixed GMSB parameters are as in
\figref{omegachi}.
\label{fig:gravBBN}
}
\end{figure}

So far, we have considered constraints from particle physics and
$\OmegaDM$.  We now turn to astrophysical constraints.  In the
preferred band, the gravitino is light and dominantly couples through
its Goldstino components. The neutralino decay widths are $\Gamma(\chi
\to \gamma \gravitino) = ( \cos^2 \theta_W / 48\pi ) (m_{\chi}^5 /
\mgravitino^2 M_*^2)$ and $\Gamma(\chi \to Z \gravitino) = ( \sin^2
\theta_W / 48\pi ) ( m_{\chi}^5 / \mgravitino^2 M_*^2) \left[ 1 -
\left( m_Z^2 / m_{\chi}^2 \right) \right]^4$.  As shown in
\figref{gravBBN}, these imply lifetimes $\tau \agt 0.01~\s$ in the
preferred band.  Such late decays are constrained by entropy
production, $\mu$ distortions of the cosmic microwave background, Big
Bang nucleosynthesis (BBN)~\cite{Feng:2003xh}, and small scale
structure~\cite{Borgani:1996ag}.  We find that the last two are most
stringent, and so focus on them here.

Standard BBN agrees reasonably well with observations.  This agreement
constrains electromagnetic (EM) and hadronic energy release in late
decays, which may be parameterized by $\xi_{\text{i}} \equiv
\epsilon_{\text{i}} B_{\text{i}} Y_{\chi}$, where $\text{i} =
\text{EM}, \text{had}$, $\epsilon_{\text{i}}$ is the EM/hadronic
energy released in each neutralino decay, $B_{\text{i}}$ is the
branching fraction into EM/hadronic components, and $Y_{\chi} \equiv
n_{\chi}/n_{\gamma}^{\text{BG}}$, where $n_{\gamma}^{\text{BG}} = 2
\zeta(3) T^3 / \pi^2$.  We have determined the $\xi_{\text{i}}$
following the prescription of Refs.~\cite{Feng:2004mt} and compared
them to the constraints given in Ref.~\cite{Jedamzik:2006xz}. The BBN
constraints are shown in \figref{gravBBN} and are stringent --- in
this scenario, neutralinos are long-lived and greatly overproduced,
resulting in large energy release. In the region of parameter space
with $0.097 < \Omega_{\gravitino} h^2 < 0.113$, the EM (hadronic)
constraint requires lifetimes $\tau \alt 10^5~\s\ (0.1~\s)$ and
$m_{\chi} \agt 200~\gev \ (1~\tev)$.

Dark matter produced in late decays also may suppress structure on
small scales~\cite{Borgani:1996ag}.  The free-streaming scale
$\lambdafs = \int^{t_{\text{EQ}}}_{\tau} [v(t)/a(t)] dt$ is well
approximated by
\begin{equation}
\lambda_{\text{FS}} \simeq
1.0~\Mpc \left[\frac{u_{\tau}^2 \tau}{10^6~\s}\right]^{\frac{1}{2}}
\left[ 1-0.07 \ln \left(\frac{u_{\tau}^2 \tau}{10^6~\s} \right) 
\right] 
\end{equation}
in the present context, where $u_{\tau} \equiv |\vec{p}_{\gravitino}|
/ m_{\gravitino}$ at decay time $\tau$, and we have neglected the
effect of $m_Z$ on kinematics and other small effects.  Values of
$\lambdafs$ are given in \figref{gravBBN}; they are essentially
independent of $m_{\gravitino}$.  Current
constraints~\cite{Viel:2006kd} require $\lambdafs \alt 0.2~\Mpc$, but
values near this bound may be preferred by observations.  Remarkably,
constraints from small scale structure are satisfied in the region of
parameter space allowed by BBN, flavor and CP bounds, but just barely
--- Goldilocks supersymmetry therefore predicts ``warm'' dark matter
and may explain the suppression of power on scales $\sim 0.1~\Mpc$.

In summary, we have proposed a simple model in which the flavor and
overdensity problems of neutralino dark matter are simultaneously
solved.  In the specific framework considered here, the preferred
model is high-scale GMSB, with $m_{\gravitino} \sim 1~\gev$, $\sqrt{F}
\sim 10^9~\gev$, $\Omegachi \sim 100$, and $m_{\chi} \sim 2~\tev$.
This last mass scale is unnaturally high, but is dictated by EDM
constraints, irrespective of cosmology.  More generally, this scenario
de-emphasizes ``cosmologically preferred'' models with $\Omegachi \sim
0.1$, and implies that models typically considered excluded by
neutralino overclosure may, in fact, be viable and preferred.

{\em Acknowledgments} --- We thank Eva Silverstein for stimulating
conversations in early stages of this work.  JLF is supported in part
by NSF grants PHY--0239817 and PHY--0653656, NASA grant NNG05GG44G,
and the Alfred P.~Sloan Foundation.  BTS is supported in part by NSF
grant PHY--0239817. FT is supported in part by NSF grant PHY--0355005.



\end{document}